\documentclass[aps,prl,twocolumn]{revtex4}
\usepackage{graphicx}

\newcommand\beq{\begin{equation}}
\newcommand\eeq{\end{equation}}

\def\d{\partial}

\def\nm{\rm nm} 
\def\mum{\mu {\rm m}} 
\def\ps{\rm ps} 

\begin{document}
\title{Generation and remote detection of THz sound using semiconductor superlattices}
\author{M. Trigo}
\email{mtrigo@umich.edu}
\author{T. A. Eckhause}
\author{J. K. Wahlstrand}
\altaffiliation[Currently at ]{JILA-NIST and University of Colorado, Boulder, CO}
\author{R. Merlin}
\affiliation{Department of Physics, University of Michigan, Ann Arbor, MI 48109-1040}

\author{M. Reason and R. S. Goldman}
\affiliation{Department of Materials Science and Engineering, University of Michigan, Ann Arbor, MI 48109-2136}

\begin{abstract}
The authors introduce a novel approach to study the propagation of high frequency acoustic phonons in which the generation and detection involves two spatially separated superlattices $\sim~1~{\rm \mu m}$ apart. Propagating modes of frequencies up to $\sim 1~{\rm THz}$ escape from the superlattice where they are generated and reach the second superlattice where they are detected. The measured frequency spectrum reveals finite size effects, which can be accounted for by a continuum elastic model.
\end{abstract}

\maketitle
With the availability of ultrafast laser pulses, there has been much effort in the study of the propagation of high frequency acoustic modes using picosecond ultrasonics\cite{thomsen1986,grahn1988}. In this approach, the absorption length of the light determines the wavelength range of the  acoustic waves that are generated.  Alternatively, high frequency ultrasound can be excited by light pulses using periodic structures such as a superlattice (SL)\cite{basseras1994,chern2003,yamamoto1994,bartels1998,ozgur2001,merlin1997,mizoguchi1999}. In such structures, the modulation of the elastic properties introduces an additional periodicity that backfolds the bulk phonon branches into a smaller Brillouin zone\cite{colvard1980}. The new zone contains modes that have nonzero frequency at the zone-center and can couple to light. In contrast to the picosecond ultrasonics method, the wavelength of the waves generated is determined by the period of the SL.
Propagating folded acoustic phonons have been observed using two-color pump-probe experiments where the pump and probe are absorbed in different regions inside the sample\cite{mizoguchi1999}.
In this work, we demonstrate a new method to probe acoustic vibrations using two identical SLs as acoustic transducers. The acoustic waves are generated and detected by ultrafast laser pulses in optical pump-probe experiments. We identify the modes observed with the aid of a continuum model, which  also explains a fine structure that arises due to the finite size of the SLs.

The time resolved studies were performed at $77~{\rm K}$ using a standard pump-probe setup, in which an ultrafast optical pulse is used to coherently excite acoustic vibrations. The probe pulses reflect off the sample, and we record the change in the reflectivity due to the presence of the coherent phonons as a function of the time delay between the pump and probe pulses.
We use $\sim 70~{\rm fs}$ pulses from an  optical parametric amplifier tunable in the visible range ($400 - 650~{\rm nm}$). The energy of the pulses is $\sim 80~{\rm nJ}$ and the pulse wavelength was centered at $530~{\rm nm}$.
The probe can be focused on the same side of the sample as the pump, or on the opposite side; see Fig. \ref{gaas_doubleSL}.

We studied two samples grown by molecular beam epitaxy on a (100) GaAs substrate that consist of a thick GaAs layer of $1.2~\mum$ or $0.6~\mum$  between two identical superlattices of 25 periods of $12~{\rm \AA}$-thick GaAs/$34~{\rm \AA}$-thick AlAs. In what follows, we refer to these samples as sample $A$ and $B$, respectively.  To access the second SL we removed the substrate by mechanical polishing and chemical etching.
Our scheme makes use of the photoelastic coupling of acoustic vibrations to light through the zone folding of the acoustic branches. The absorption of the pump pulse in one of the SLs launches a high frequency acoustic wave that propagates towards the second SL where it is detected by the delayed probe pulse (Fig. \ref{gaas_doubleSL}). This particular design makes it possible to generate and detect vibrations in spatially separated regions.

The samples were characterized by Raman spectroscopy. Spectra obtained with the $514.5~\nm$ line of an Argon-ion laser in the backscattering geometry show a doublet near $\sim 1~{\rm THz}$ which corresponds to the first zone-center folded acoustic modes\cite{jusserand_lss}. The frequencies obtained from these experiments match the calculated  ones using an infinite continuum elastic model, originally developed by Rytov \cite{rytov1956}.

In Fig. \ref{time_traces} we show experimental differential reflectivity data for the two samples. In (a) we show a typical time trace for the configuration in which the pump and the probe arrive at the same face of the sample. The data clearly exhibit a high frequency oscillation superimposed on top of a large, lower frequency vibration due to stimulated Brillouin scattering\cite{thomsen1986}.
Results for the situation in which the pump and the probe strike on opposite sides of our structure are shown in Fig. \ref{time_traces} (b). For samples $A$ and $B$ we observe, respectively, an oscillating transient in the reflectivity arriving after time delays of $\sim 115~\ps$ and $\sim230~\ps$. These values agree well with the time-of-flight of  acoustic waves traveling across the bulk layer. We note that the high frequency folded phonons  arrive at the second SL after the transit time, as shown in the inset of Fig. \ref{time_traces} (b), which is an expanded view of the region indicated by the dotted box.
The amplitude of the oscillation due to the folded phonons is nearly the same in the data of Fig. \ref{time_traces} (a) and Fig. \ref{time_traces} (b), indicating that they propagate without a measurable attenuation through the GaAs layer.

Figure \ref{fftSL} (a) shows the Fourier transform of the trace presented in the inset of Fig. \ref{time_traces} (b). There are four main features present, a large amplitude vibration due to Brillouin scattering\cite{thomsen1986} at $70~{\rm GHz}$ (not shown in the figure) and a triplet due to zone-center folded phonons with components near $1~{\rm THz}$. Two modes of the triplet correspond to the back-scattering wavevector (BS) while the remaining one (FS) originates from the zone center mode that is Raman active \cite{jusserand_lss}.
The frequency of the BS modes is determined by the wavelength of the probe pulse. This dependence can be understood from the following argument. The change in reflectivity can be written as \cite{thomsen1986}
\begin{equation}\label{probe_rqs}
\Delta r \propto \int_{-\infty}^{+\infty} e^{2 i k_p z}  P(z) \eta(z,t) d z,
\end{equation}
where $\eta(z,t) = \partial u(z,t)/\partial z$ is the time dependent strain, $P(z)$ is the corresponding component of the photoelastic tensor and $k_p$ is the wavevector of the probe.
In an infinite structure, Eq. (\ref{probe_rqs}) has non-zero contributions from those components of $\eta$ that satisfy the phase matching condition $q=\pm 2 k_p$. This means that, regardless of the generation mechanism, the probe selects only those modes that satisfy wavevector conservation. Furthermore, the propagating strain pulse moves at the speed of sound and, therefore, the corresponding frequency for the phase matched low frequency oscillations in Fig. \ref{time_traces} is  $\nu_{\rm BR} = k_p/\pi v$ where $v$ is the speed of sound of the SL. For the same reasons, the folded phonon doublet at $\sim 1~{\rm THz}$ is composed of modes which, in the extended zone scheme, have wavevectors $q = 2\pi/D \pm 2 k_p$, where $D$ is the SL period.

We calculated the change in reflectivity induced by the pump-generated acoustic vibrations. The corresponding equation of motion is \cite{thomsen1986}
\beq \label{main_wave} \rho(z) \frac{\d^2 u(z,t)}{\d t^2} = 
\frac{\d \sigma(z,t)}{\d z}
\eeq
where $\rho$ is the density, $\sigma(z,t) = C(z) \d u(z,t)/\d z$ is the stress, $C(z)$ is the elastic stiffness, and $u(z,t)$ is the atomic displacement. The energy deposited by the laser pulse sets up an initial stress  $\sigma_0(z) = K(z) |E(z)|^2$, where $E(z)$ is the electric field of the pump pulse and $K(z)$ is a square wave that represents the absorption in different layers of the SL \cite{grahn1988}.
The use of a continuum model for structures with such small periods is somewhat arguable since the bulk properties of each constituent are not well defined and, hence, the local index of refraction or, equivalently, the absorbed energy in each layer is not known. With this in mind, we adjusted the ratio $K_{\rm GaAs}/K_{\rm AlAs}$ to reproduce the relative intensities of the observed modes.
Regardless of the generation mechanism, the waves are detected by stimulated light scattering through the photoelastic mechanism, as stated in Eq. (\ref{probe_rqs}).

Calculated spectra are shown in  Figs. \ref{fftSL} (b) and (c) together with the  experimental result (a). The best fit was obtained for $K_{\rm GaAs}/K_{\rm AlAs} = 1.2$. For clarity, we have included the dispersion relation for the infinite SL (lower panel). In the spectrum corresponding to a single \emph{semi-infinite} SL, Fig. \ref{fftSL} (c), the flat features with zero intensity at $0.5~{\rm THz}$ and $1~{\rm THz}$ reflect the acoustic minigaps whereas the  doublet at $\sim 1~{\rm THz}$ corresponds to the folded phonons observed in  back-scattering Raman spectra \cite{colvard1980}.

Figure \ref{fftSL} (b) shows the calculated spectrum for the finite SL in the case where the probe is incident on the second SL. The calculations exhibit a complex structure with an overall modulation of the intensity, which is also visible on the lower side of the measured spectrum in (a).
The frequency spacing of the comb-like structure is $\Delta \nu \sim 20~{\rm GHz}$. This value corresponds to the inverse of the transit time of the acoustic pulse through the SL. This periodic modulation manifests itself in the time domain traces as a time delayed signal. We note that the high frequency oscillations shown in the inset of Fig. \ref{time_traces} are delayed from the main Brillouin oscillations. This delay, of $\tau \sim 50~{\rm ps}$, is consistent with the frequency spacing of the comb.

The folded phonons in Fig. \ref{fftSL} (b) also show a fine structure which arises due to the finite size of the SL. In the experimental data, this fine structure is particularly evident in the higher frequency BS-peak.
In addition, the peak at the zone center (FS) is also visible. This feature appears stronger in the experiment than in the calculations. We tentatively attribute this discrepancy to the interface contribution to $\Delta r$\cite{merlin1997}, which is not taken into account in Eq. (\ref{probe_rqs}).

In conclusion, we introduced a new method to probe the dynamics of acoustic phonons of frequencies in the THz range using two spatially separated SLs. The frequency spectrum of the detected waves has a complex structure that originates in the finite number of periods of the SLs. A continuum elastic model accurately predicts the observed frequencies as well as the fine structure. However, some parameters in the model had to be adjusted to obtain the correct amplitudes in the spectrum.

This work was supported by the NSF FOCUS Physics Frontier Center.


\begin{thebibliography}{12}
\expandafter\ifx\csname natexlab\endcsname\relax\def\natexlab#1{#1}\fi
\expandafter\ifx\csname bibnamefont\endcsname\relax
  \def\bibnamefont#1{#1}\fi
\expandafter\ifx\csname bibfnamefont\endcsname\relax
  \def\bibfnamefont#1{#1}\fi
\expandafter\ifx\csname citenamefont\endcsname\relax
  \def\citenamefont#1{#1}\fi
\expandafter\ifx\csname url\endcsname\relax
  \def\url#1{\texttt{#1}}\fi
\expandafter\ifx\csname urlprefix\endcsname\relax\def\urlprefix{URL }\fi
\providecommand{\bibinfo}[2]{#2}
\providecommand{\eprint}[2][]{\url{#2}}

\bibitem[{\citenamefont{{C. Thomsen} et~al.}(1986)\citenamefont{{C. Thomsen},
  {H. T. Grahn}, {H. J. Maris}, and {J. Tauc}}}]{thomsen1986}
\bibinfo{author}{\bibnamefont{{C. Thomsen}}}, \bibinfo{author}{\bibnamefont{{H.
  T. Grahn}}}, \bibinfo{author}{\bibnamefont{{H. J. Maris}}}, \bibnamefont{and}
  \bibinfo{author}{\bibnamefont{{J. Tauc}}}, \bibinfo{journal}{Phys. Rev. B}
  \textbf{\bibinfo{volume}{34}}, \bibinfo{pages}{4129} (\bibinfo{year}{1986}).

\bibitem[{\citenamefont{{H. T. Grahn} et~al.}(1988)\citenamefont{{H. T. Grahn},
  {H. J. Maris}, {J. Tauc}, and {B. Abeles}}}]{grahn1988}
\bibinfo{author}{\bibnamefont{{H. T. Grahn}}},
  \bibinfo{author}{\bibnamefont{{H. J. Maris}}},
  \bibinfo{author}{\bibnamefont{{J. Tauc}}}, \bibnamefont{and}
  \bibinfo{author}{\bibnamefont{{B. Abeles}}}, \bibinfo{journal}{Phys. Rev. B}
  \textbf{\bibinfo{volume}{38}}, \bibinfo{pages}{6066} (\bibinfo{year}{1988}).

\bibitem[{\citenamefont{{P. Bass\'eras} et~al.}(1994)\citenamefont{{P.
  Bass\'eras}, {S. M. Gracewski}, {G. W. Wicks}, and {R. J. D.
  Miller}}}]{basseras1994}
\bibinfo{author}{\bibnamefont{{P. Bass\'eras}}},
  \bibinfo{author}{\bibnamefont{{S. M. Gracewski}}},
  \bibinfo{author}{\bibnamefont{{G. W. Wicks}}}, \bibnamefont{and}
  \bibinfo{author}{\bibnamefont{{R. J. D. Miller}}}, \bibinfo{journal}{J. Appl.
  Phys.} \textbf{\bibinfo{volume}{75}}, \bibinfo{pages}{2761}
  (\bibinfo{year}{1994}).

\bibitem[{\citenamefont{{Gia-Wei Chern} et~al.}(2003)\citenamefont{{Gia-Wei
  Chern}, {Kung-Hsuan Lin}, {Yue-Kai Huang}, and {Chi-Kuang Sun}}}]{chern2003}
\bibinfo{author}{\bibnamefont{{Gia-Wei Chern}}},
  \bibinfo{author}{\bibnamefont{{Kung-Hsuan Lin}}},
  \bibinfo{author}{\bibnamefont{{Yue-Kai Huang}}}, \bibnamefont{and}
  \bibinfo{author}{\bibnamefont{{Chi-Kuang Sun}}}, \bibinfo{journal}{Phys. Rev.
  B} \textbf{\bibinfo{volume}{67}}, \bibinfo{pages}{121303}
  (\bibinfo{year}{2003}).

\bibitem[{\citenamefont{{A. Yamamoto} et~al.}(1994)\citenamefont{{A. Yamamoto},
  {T. Mishina}, {Y. Masumoto}, and {M. Nakayama}}}]{yamamoto1994}
\bibinfo{author}{\bibnamefont{{A. Yamamoto}}},
  \bibinfo{author}{\bibnamefont{{T. Mishina}}},
  \bibinfo{author}{\bibnamefont{{Y. Masumoto}}}, \bibnamefont{and}
  \bibinfo{author}{\bibnamefont{{M. Nakayama}}}, \bibinfo{journal}{Phys. Rev.
  Lett.} \textbf{\bibinfo{volume}{73}}, \bibinfo{pages}{740}
  (\bibinfo{year}{1994}).

\bibitem[{\citenamefont{{A. Bartels} et~al.}(1998)\citenamefont{{A. Bartels},
  {T. Dekorsy}, {H. Kurz}, and {K. K\"ohler}}}]{bartels1998}
\bibinfo{author}{\bibnamefont{{A. Bartels}}}, \bibinfo{author}{\bibnamefont{{T.
  Dekorsy}}}, \bibinfo{author}{\bibnamefont{{H. Kurz}}}, \bibnamefont{and}
  \bibinfo{author}{\bibnamefont{{K. K\"ohler}}}, \bibinfo{journal}{Appl. Phys.
  Lett.} \textbf{\bibinfo{volume}{72}}, \bibinfo{pages}{2844}
  (\bibinfo{year}{1998}).

\bibitem[{\citenamefont{{\"{U}. \"{O}zg\"{u}r}
  et~al.}(2001)\citenamefont{{\"{U}. \"{O}zg\"{u}r}, {C. Lee}, and {H. O.
  Everitt}}}]{ozgur2001}
\bibinfo{author}{\bibnamefont{{\"{U}. \"{O}zg\"{u}r}}},
  \bibinfo{author}{\bibnamefont{{C. Lee}}}, \bibnamefont{and}
  \bibinfo{author}{\bibnamefont{{H. O. Everitt}}}, \bibinfo{journal}{Phys. Rev.
  Lett.} \textbf{\bibinfo{volume}{86}}, \bibinfo{pages}{5604}
  (\bibinfo{year}{2001}).

\bibitem[{\citenamefont{{K. Mizoguchi} et~al.}(1999)\citenamefont{{K.
  Mizoguchi}, {M. Hase}, {S. Nakashima}, and {M. Nakayama}}}]{mizoguchi1999}
\bibinfo{author}{\bibnamefont{{K. Mizoguchi}}},
  \bibinfo{author}{\bibnamefont{{M. Hase}}}, \bibinfo{author}{\bibnamefont{{S.
  Nakashima}}}, \bibnamefont{and} \bibinfo{author}{\bibnamefont{{M.
  Nakayama}}}, \bibinfo{journal}{Phys. Rev. B} \textbf{\bibinfo{volume}{60}},
  \bibinfo{pages}{8262} (\bibinfo{year}{1999}).

\bibitem[{\citenamefont{Merlin}(1997)}]{merlin1997}
\bibinfo{author}{\bibfnamefont{R.}~\bibnamefont{Merlin}},
  \bibinfo{journal}{Solid State Commun.} \textbf{\bibinfo{volume}{102}},
  \bibinfo{pages}{207} (\bibinfo{year}{1997}).

\bibitem[{\citenamefont{{C. Colvard} et~al.}(1980)\citenamefont{{C. Colvard},
  {R. Merlin}, {M. V. Klein}, and {A. C. Gossard}}}]{colvard1980}
\bibinfo{author}{\bibnamefont{{C. Colvard}}}, \bibinfo{author}{\bibnamefont{{R.
  Merlin}}}, \bibinfo{author}{\bibnamefont{{M. V. Klein}}}, \bibnamefont{and}
  \bibinfo{author}{\bibnamefont{{A. C. Gossard}}}, \bibinfo{journal}{Phys. Rev.
  Lett.} \textbf{\bibinfo{volume}{45}}, \bibinfo{pages}{298}
  (\bibinfo{year}{1980}).

\bibitem[{\citenamefont{{B. Jusserand} and {M. Cardona}}(1989)}]{jusserand_lss}
\bibinfo{author}{\bibnamefont{{B. Jusserand}}} \bibnamefont{and}
  \bibinfo{author}{\bibnamefont{{M. Cardona}}}, in
  \emph{\bibinfo{booktitle}{Light Scattering in Solids}}, edited by
  \bibinfo{editor}{\bibfnamefont{M.}~\bibnamefont{Cardona}} \bibnamefont{and}
  \bibinfo{editor}{\bibfnamefont{G.}~\bibnamefont{G\"{u}ntherodt}}
  (\bibinfo{publisher}{Springer}, \bibinfo{address}{Berlin},
  \bibinfo{year}{1989}), p.~\bibinfo{pages}{49}.

\bibitem[{\citenamefont{{S. M. Rytov}}(1956)}]{rytov1956}
\bibinfo{author}{\bibnamefont{{S. M. Rytov}}}, \bibinfo{journal}{Akust. Zh.}
  \textbf{\bibinfo{volume}{2}}, \bibinfo{pages}{71} (\bibinfo{year}{1956}).

\end{thebibliography}
\newpage

\begin{figure}[htb]
	\caption{Schematic diagram of the double SL structure and the geometry of the pump-probe experiments. The samples consist of a thick GaAs layer of $1.2~\mum$ or $0.6~\mum$ between two identical SLs.\label{gaas_doubleSL}}
\end{figure}

\begin{figure}
	\caption{
	Time-domain data for samples A and B. Pump pulses generate sound waves on the front superlattice, that are later detected on the front (a) and the backside (b) of the sample. Trace (a) and the inset of (b), obtained on sample A, show folded-phonon oscillations superimposed on the larger, low-frequency oscillations associated with stimulated Brillouin scattering.\label{time_traces}}
\end{figure}

\begin{figure}
	\caption{(a) Fourier transform of the time trace for sample A. We observe three peaks that correspond to the backscattering (BS) and forward scattering (FS) modes accessible in Raman experiments; (b) calculated spectrum for the double SL structure, where the probe is incident on the second SL and (c) calculated spectrum for a semi-infinite structure showing the BS doublet. The lower panel shows the dispersion relation for an infinite SL.\label{fftSL}}
\end{figure}

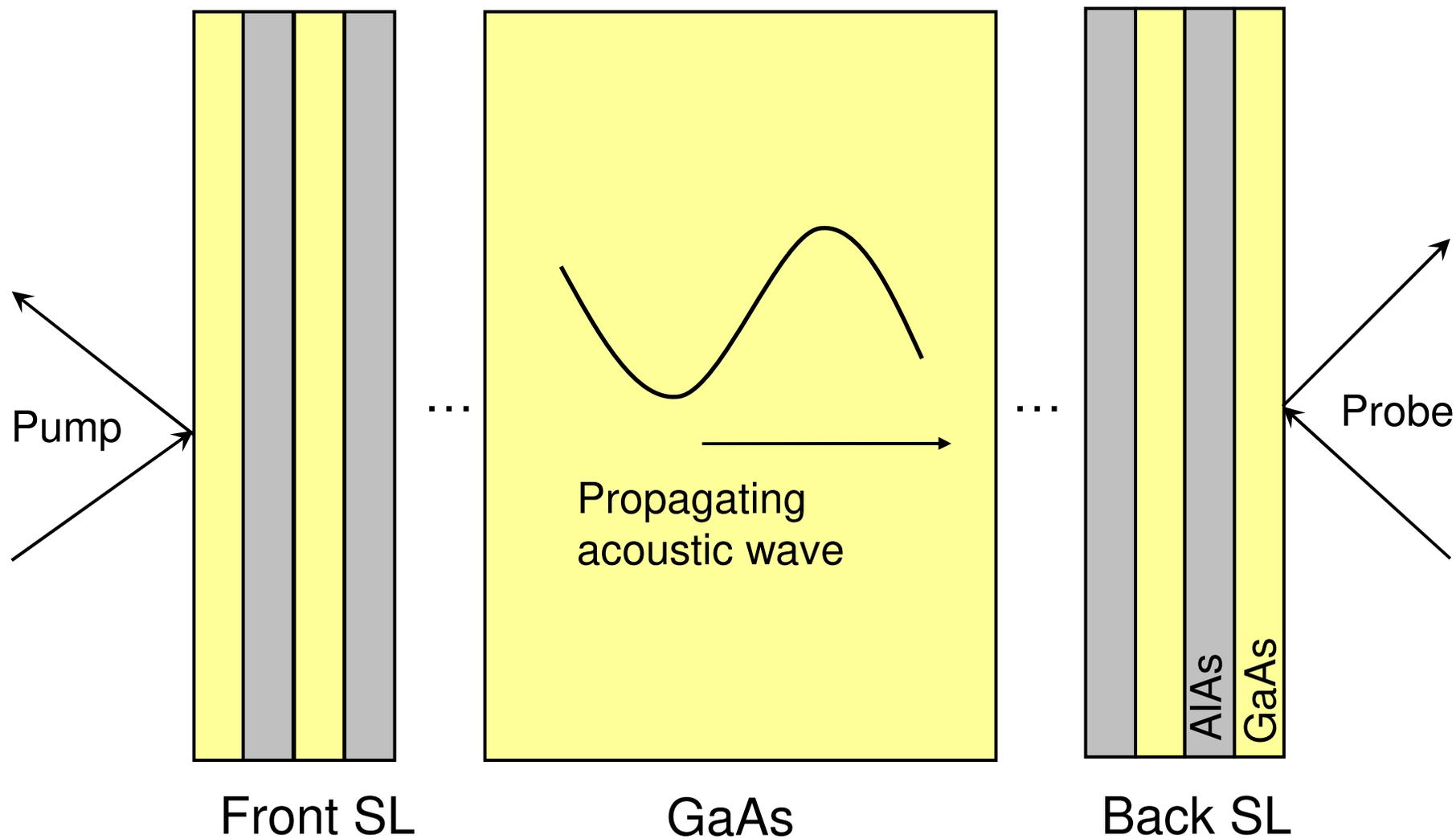
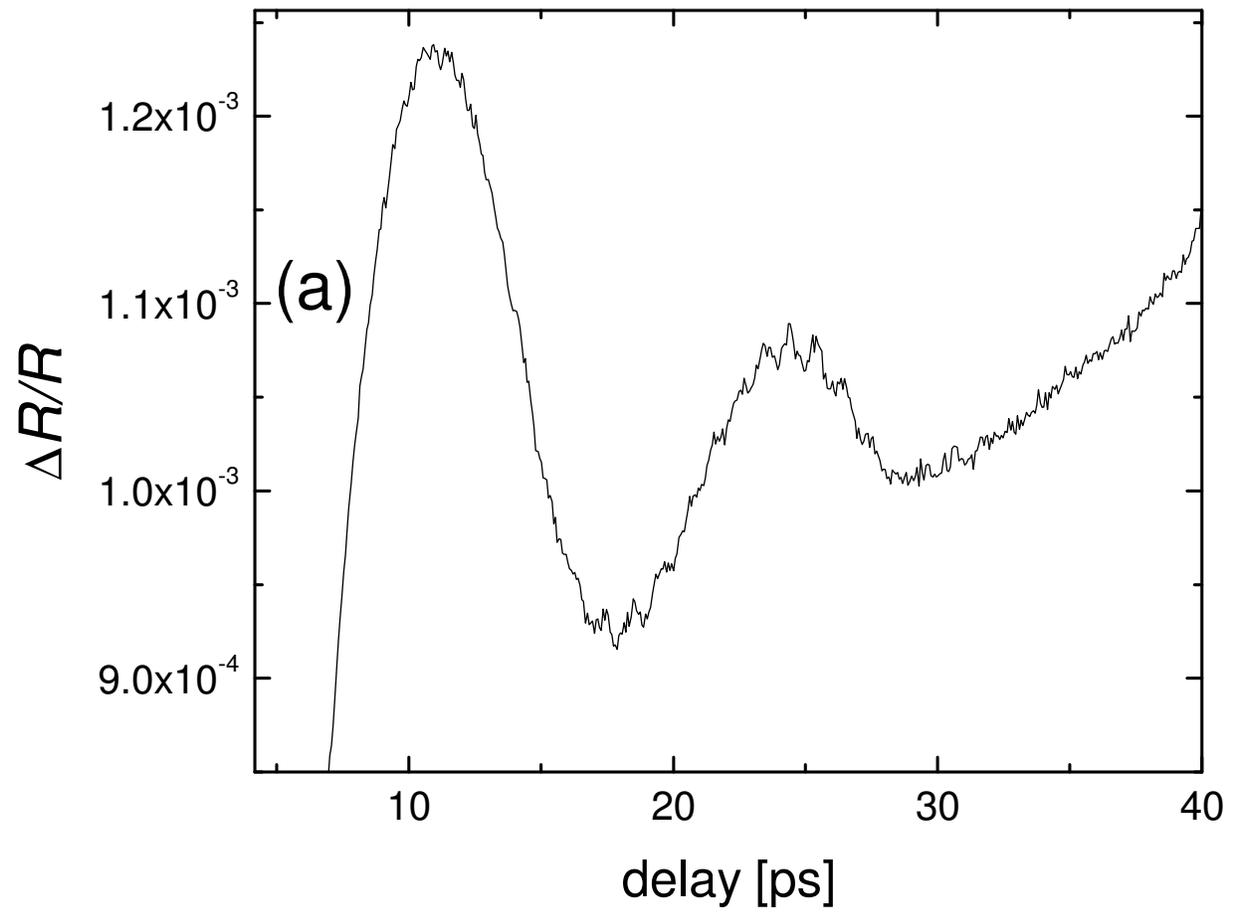
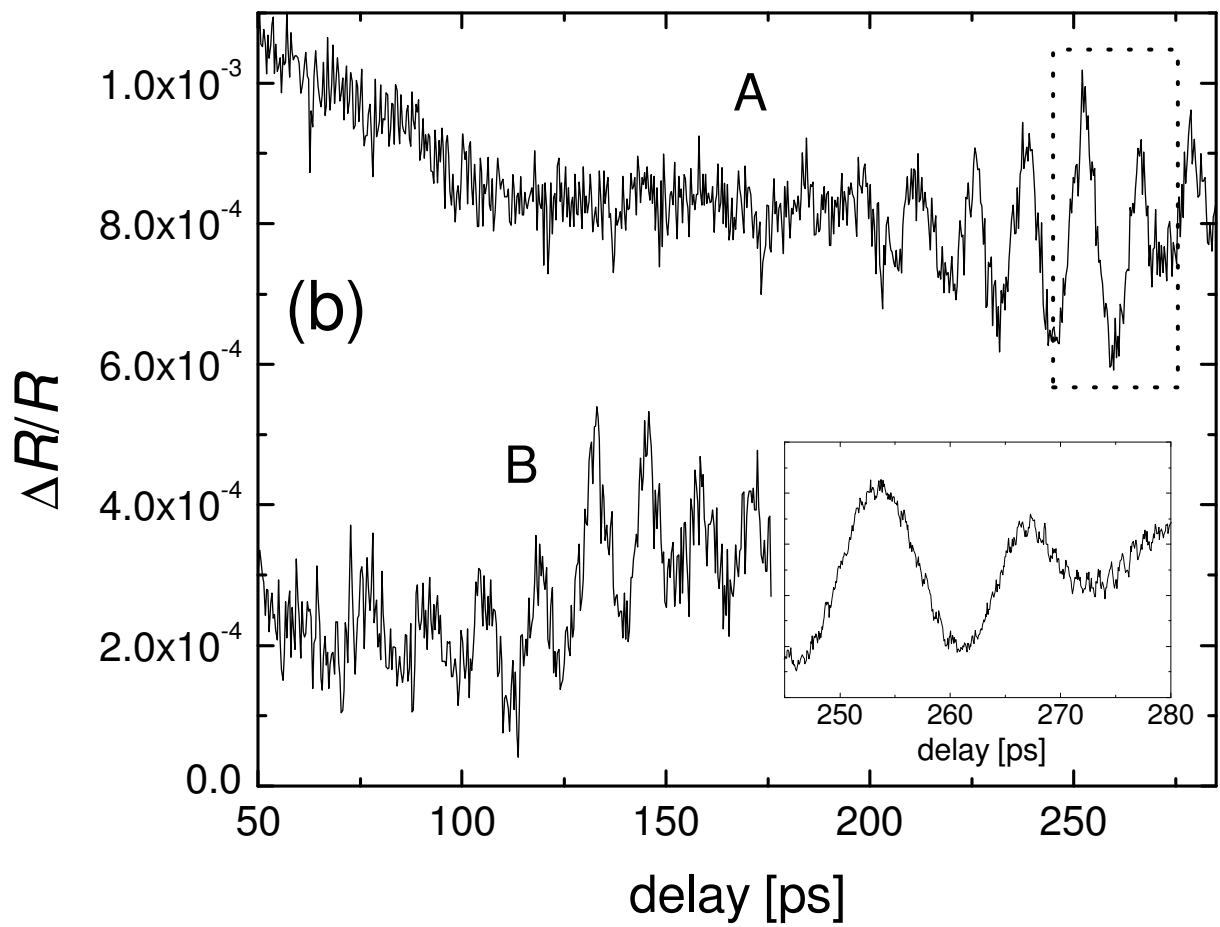
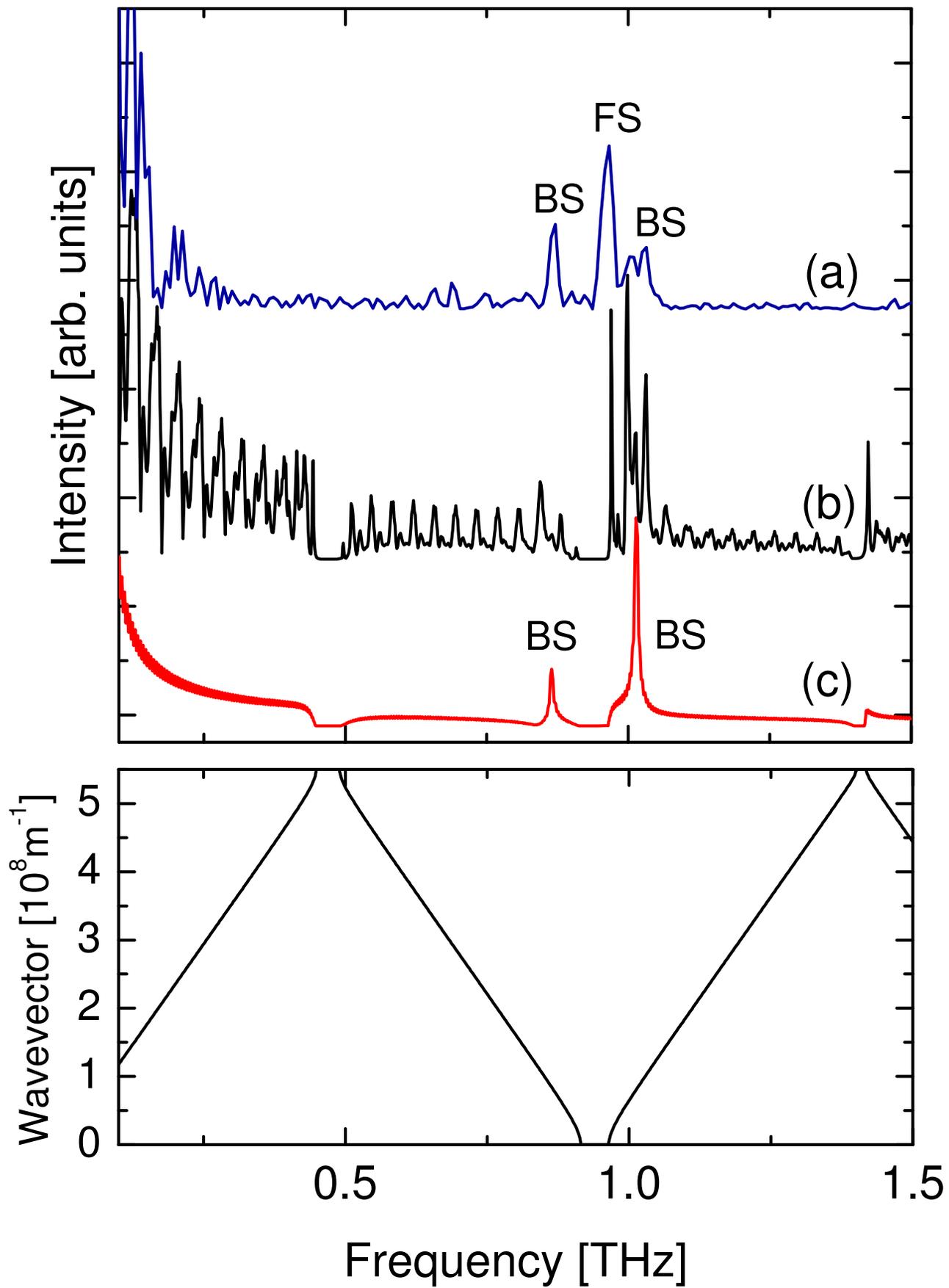
\end{document}